\documentclass[preprint]{aastex631}
\usepackage[T1]{fontenc}
\usepackage{mathtools}
\usepackage{afterpage}

\shorttitle{LivRed Rotation-Age Relationships}
\shortauthors{Engle \& Guinan}

\graphicspath{{./}{figures/}}

\begin{document}

\title{Living with a Red Dwarf: The Rotation--Age Relationships of M Dwarfs}

\author[0000-0001-9296-3477]{Scott G. Engle}

\author[0000-0002-4263-2650]{Edward F. Guinan}
\affiliation{Villanova University \\
Dept. of Astrophysics and Planetary Science \\
800 E. Lancaster Ave \\
Villanova, PA 19085, USA}

\begin{abstract}

Age is a fundamental stellar property, yet for many stars it is difficult to
reliably determine. For M dwarfs it has been notoriously so. Due to their 
lower masses, core hydrogen fusion proceeds at a much slower rate in M dwarfs
than it does in more massive stars like the Sun. As a consequence, more
customary age determination methods (e.g. isochrones and asteroseismology) are
unreliable for M dwarfs. As these methods are unavailable, many have searched
for reliable alternatives. M dwarfs comprise the overwhelming majority of the
nearby stellar inventory, which makes the determination of their fundamental
parameters even more important. Further, an ever-increasing number of exoplanets
are being found to orbit M dwarfs and recent studies have suggested they may
relatively higher number of low-mass planets than other spectral types. Determining
the ages of M dwarfs then allows us to better study any hosted exoplanets, as well.
Fortunately, M dwarfs possess magnetic activity and stellar winds like
other cool dwarf stars. This causes them to undergo the spindown effect (rotate
with longer periods) as they age. For this reason, stellar rotation rate has
been considered a potentially powerful age determination parameter for over 50 years.
Calibrating reliable age-rotation relationships for M dwarfs has been a lengthy
process, but here we present the age-rotation relationships for $\sim$M0--6.5 dwarfs,
determined as part of the \textit{Living with a Red Dwarf} program. These
relationships should prove invaluable for a wide range of stellar astrophysics
and exoplanetary science applications.

\end{abstract}


\section{Introduction \& Background: Studying M Dwarfs} \label{sec:intro}

Main sequence (dwarf) M stars (dM stars; red dwarfs; referred to as
\textit{M dwarfs} hereafter) represent the cool, low mass, low luminosity end of the main
sequence, and comprise $\sim$75\% of all stars in the solar neighborhood
\citep{2021AA...650A.201R}. This study specifically focuses on M0 V --
$\sim$M6.5 V stars, with properties ranging
from: Mass $\approx$ 0.6 -- 0.1 $M_\odot$; Radius $\approx$ 0.6 -- 0.1
$R_\odot$; Luminosity $\approx$ 0.06 – 0.001 $L_\odot$ and temperatures
\textit{T}$_{\rm eff}$ = 3900 -- 2850 K\footnote{\url{https://www.pas.rochester.edu/~emamajek/EEM_dwarf_UBVIJHK_colors_Teff.txt}}. 

M dwarfs have received substantial attention during the 2000’s, prompted in
part by the discovery that these numerous stars host a relatively large number
of terrestrial-size planets \citep{2023arXiv230103442R,2020AJ....160..237F,2020MNRAS.498.2249H} when compared to stars of higher mass. Aside
from the large number of nearby M dwarfs available for study, they also make
very attractive targets for terrestrial planet searches and research programs
as such planets are more readily detected through radial velocity motions and
planetary transits due to the low masses and small radii of the M dwarf host
stars. Estimates of the frequency of potentially habitable planets (PHP)
hosted by M dwarfs have been made primarily from \textit{Kepler} Mission data, but also
from numerous radial velocity studies. Conservative estimates place the
planetary frequency around 15\% \citep{2013ApJ...767...95D} and studies including expanded
circumstellar Habitable Zone (HZ) estimates indicate higher frequencies of
$\sim$30--40\% \citep{2020MNRAS.498.2249H,2013ApJ...767L...8K}. If a slightly conservative `middle ground' of 25\% is adopted,
it implies that within 10 pc ($\sim$33 ly) of the Sun (a volume of space containing
$\sim$240 M dwarfs), there should be $\sim$60 potentially habitable Earth-size
planets. Extrapolating to include the entire Milky Way raises the possibility that
billions of Earth-size planets are orbiting within the habitable zones of M dwarfs. 

M dwarfs are equally fascinating targets for stellar astrophysics. Relative to
their sizes, they display enhanced magnetic dynamo activity due to the extent of
their interior convective zones. As a result, their coronal and
chromospheric emissions are also relatively strong, compared to their bolometric
luminosities. They have comparatively slow core nuclear reaction rates, however, 
which makes them rather `fuel efficient' and results in their long main-sequence lifetimes.
More massive M dwarfs can live on the
main sequence for over 100 Gyr while those of lower mass (M $<$ 0.2 $M_\odot$)
can live as long as $\sim$1 trillion ($\sim$10$^{12}$) years \citep{2016ApJ...823..102C}. Due to this,
no M dwarfs have yet to evolve off the main sequence. Another consequence of
their long lifetimes, however, is that once M dwarfs reach the core hydrogen-fusing
main-sequence their basic physical properties (\textit{L}, \textit{T}$_{\rm eff}$,
\textit{R}) remain essentially constant over cosmological time scales (i.e.,
$\sim$14 Gyr). 

Their large numbers, longevities, and near-constant main sequence luminosities make
M dwarfs very compelling targets for programs searching for life in the universe
since, unlike our Sun, the HZs and thus exoplanet bolometric irradiances (and
planetary instellations) remain stable for tens of Gyrs or longer. However, the
stars' very slow nuclear evolution makes determining accurate stellar ages extremely
challenging \citep[see][and references therein]{2010ARAA..48..581S}. 

Fortunately, it has been known for 50 years \citep{1972ApJ...171..565S} that cool dwarfs undergo
a `spindown effect' whereby their rotation periods lengthen as they age. Since that
time, numerous studies have shown the potential that stellar rotation holds as an age
determinant -- the method known as ``gyrochronology'' \citep{2003ApJ...586..464B,2007ApJ...669.1167B,2008ApJ...687.1264M,
2011ASPC..451..285E,2018RNAAS...2...34E,2022ApJ...936..109P}. Late-F, G, K, and M dwarfs have stellar winds and magnetic
fields which act in tandem to propagate the spin down effect. The winds of these
stars are magnetically-threaded, continually carrying small amounts of each star's
mass out into space while it is still (over a certain distance) tethered to the star
itself by the magnetic field. The mass eventually escapes the magnetic field entirely,
but its magnetically-threaded tenure has already caused a slowing of the star's
rotation due to conservation of angular momentum \citep{1988ApJ...333..236K}. 

This spindown effect allows magnetic activity levels (observed through such proxies
as X-ray and UV [X--UV] emissions, and several emission features known to exist within
optical spectra) to serve as additional age determinants for cool
dwarfs. Activity-age relationships have been also been constructed for M dwarfs, and they will be
detailed in a follow-up paper. 

The largest difficulty resided in building a representative sample of M dwarfs with a
wide range of previously known ages and then determining their rotation periods. In this
paper we present these `benchmark' objects and the rotation-age relationships of M dwarfs
determined as part of the \textit{Living with a Red Dwarf} (\textit{LivRed}) Program.

\section{Dating M dwarfs: Determining Ages for (Mostly) Ageless stars} \label{sec:floats}

Age, along with mass and composition, is one of three key factors governing a star's
current state \citet{2010ARAA..48..581S}, yet it is also one of the most difficult stellar parameters to
accurately measure. As mentioned, this is particularly true with M dwarfs, for which
other commonly applied methods (e.g., isochronal, asteroseismic) for aging
a star are unreliable \citet{2021AJ....161..189L}. Observables,
such as rotation period and X-UV activity level, are known to be age-dependent and are
often related to each other, but relating either quantity to stellar age first requires
a set of M dwarfs with known ages -- a benchmark sample.

With no currently available methods for directly determining the ages of single,
isolated M dwarfs, the sample of benchmark M dwarfs has instead been built using \textit{age
by association}. Each benchmark either has a stellar companion or belongs to a larger
group or population of stars within the galaxy. For each pairing or grouping of stars, it is the
age of the companion star or the group that can also be applied to the M dwarf since
they are assumed to have formed at the same time. 

The age by association method in this study can be divided into three categories. For
young dwarfs with ages below $\sim$2 Gyr there are several well-studied `stellar groups'
(referred to as either moving groups, clusters, or associations) available. The
ages are very reliable, but sadly do not cover nearly the range that we need. There
are a limited number of additional clusters with greater ages, but other issues exist.
For example, in the clusters NGC 752, and Ruprecht 147 \citep[ages of 1.4, 2.5,
and 2.7 Gyr -- see][and references therein]{2020AA...644A..16G,2020ApJ...904..140C}
rotations periods have only been
measured for earlier M dwarfs. A small number of HR 1614 moving group members (age $\sim$2.0
Gyr -- [60]) with rotation periods were once used, but the coherence of the moving
group itself has recently been called into question \citet{2020AA...638A.154K} which prompted their removal
as benchmarks. Traditionally, the distances of highly prized targets such as M67 and NGC
188 (ages \> 4 Gyr) prevented sufficient time-series photometry any their faint M dwarf
members. However, \citet{2022ApJ...938..118D} have recently measured rotation rates within this cluster
for stars as late as $\sim$M3, showing some of the exciting recent progress in cluster
gyrochronology measures. See Section \ref{sec:m67} for further details regarding M67.

Ages can also be assigned to stars that are members of specific galactic populations.
Additional benchmarks were selected which belong to either the Thick Disk or Halo populations
(ages of $\sim$8--11 and $\sim$10--12.5 Gyr) of the Milky Way, based
primarily on the star's \textit{UVW} galactic space motions \citep{1998ApJ...497..294L,2014AA...562A..71B}, with further support
of membership from metallicity values and velocity dispersions \citep{2018MNRAS.475.1093Y}. The advanced ages
of these populations make them important benchmarks, but more direct age estimates for
individual M dwarfs, as opposed to statistically-supported, kinematically-inferred ages,
would usually be preferred.

The final and very welcome source of benchmarks is M dwarfs that belong to common
proper motion (CPM) pairs/systems with an age-determinable companion. If the companion
is a more massive (F--G dwarf) star, then a reliable age can be determined by other,
more common (e.g., isochronal and/or asteroseismic) methods and applied to the (assumed
to be coeval) M dwarf. Systems with white dwarf (WD) companions have become
increasingly useful due to advances in determining the WD progenitor star properties
\citep[see][]{2018ApJ...866...21C} that have resulted in increasingly reliable ages. 
It is always important to note that the separation of the M dwarf from its companion is
assumed to have prevented past interactions, allowing the M dwarf to evolve as if it were
a single, isolated star. Though, for specific pairs (particularly those with small
separations is small), the possibility of past interactions may exist (see
\citealt{2022ApJ...936..109P}). However, a particular benefit these
systems have over the previously mentioned CPM pairs is that the WDs do not outshine
their M dwarf companions, which facilitates CCD photometry of the M dwarfs to search
for rotation periods. These systems provided several M dwarf targets with ages older
than 2 Gyr: an age-range that was long-awaiting additional targets. Determining rotation
periods for these older M dwarfs became a primary focus of the program. 

For this study, multiple (if available) measures of each WD companion's effective
temperature (\textit{T}$_{\rm eff}$) and surface gravity (\textit{log g}) were gathered
from the recent literature, and mean values and uncertainties were determined via $\chi^{2}$
analysis. With these values, updated ages and uncertainties were calculated using both the
\texttt{WD\_Models} package, written by Dr. Sihao Cheng, and the \texttt{wdwarfdate} package,
written by Dr. Rocio Kiman \citep{2022AJ....164...62K}. Both incorporate the latest WD
cooling models available \citet{2020ApJ...901...93B} and the initial-final mass relationship
(IFMR) of \citet{2018ApJ...866...21C}. We note that this is not the only IFMR choice
available within the \texttt{wdwarfdate} package, but it is the one we selected both for
consistency with \texttt{WD\_Models} and because recent literature shows it is still the
most often used IFMR. 

\subsection{Calculating White Dwarf Ages} \label{sec:wdages}

Using ages derived from white dwarf companions is still something of a novel technique. 
The use of mean values for the white dwarfs was deemed an important method for this
program to mitigate the uncertainties that could arise from choosing any single measure for
individual targets, especially since no single study could be found that had measured
every white dwarf we wished to analyze. To give some estimate of this effect, for all benchmarks included in
\citet{2021MNRAS.508.3877G}, ages computed using their mean properties were compared
to those computed using only the \citealt{2021MNRAS.508.3877G} values. An average age
difference of $\sim$0.8 Gyr was found, though a major contributor is LP 498-26, for which 
\citealt{2021MNRAS.508.3877G} determined a fairly low \textit{log g} value. If only
LP 498-26 is removed from the comparison, the average age difference drops to $\sim$0.5 Gyr.

Additionally, as mentioned at the end of section \ref{sec:floats}, the \texttt{wdwarfdate} package
offers more than one IFMR. We chose to use that of \citet{2018ApJ...866...21C} for reasons
mentioned, although an alternative exists \citep{2020NatAs...4.1102M} which uses newer
data and appears to show a new feature in the relationship. Again, age comparisons were
made, but this time between those determined using the two different IFMRs. An average age
determination of $\sim$0.2 Gyr was found.

\subsection{The Complicated Case of M67} \label{sec:m67}

Though plotted in Figs \ref{fig:rotagelinearize} and \ref{fig:rotagesemilog}, the data
for M67 were not included when determining our relationships. \citet{2022ApJ...938..118D}
quote an age of 4 Gyr, which is the often-quoted age within the literature. However, the
full age-range quoted in \citet{2022ApJ...938..118D} is 3.5 -- 5 Gyr, and isochronal studies
perhaps point more towards an age of $\sim$3.5 Gyr \citep[e.g.,][]{2023AA...672A.159G}. A number of
gyrochronology studies appear to arrive at or near 4 Gyr, but we felt perhaps it wouldn't be
appropriate to use a primarily gyrochronology-based age when calibrating the rotation-age
relationships presented here. 

Using the relationships and rotation values listed here, we determine an early M dwarf age
for M67 of $\sim$3.2 $\pm$ 0.8 Gyr and a mid-late M dwarf age of $\sim$3.6 $\pm$ 1.0 Gyr.

\section{Staring at M Dwarfs: Determining Rotation Periods} \label{sec:rot-act}

The surface features (e.g. starspots) of cool, main sequence stars will be brought in
and out of view as the stars rotate, if the orientation of the star (inclination of
the star's rotation axis relative to our line sight) and the star spot surface 
distribution are favorable. Repeatedly measuring stellar brightness via photometry can
determine the rotation periods by revealing cyclical changes in brightness over time.
This was the preferred method for determining benchmark rotation periods, as it is
precise and works for very long rotation periods where spectroscopic measures of rotation
velocity become ambiguous. Measuring rotation via photometry is a very straightforward
process on paper, but in practice substantial difficulties can arise when dealing with M
dwarfs. As the \textit{Living with a Red Dwarf} program was designed to study the crucial missing
age-range of $>$3 Gyr, it was unknown at the outset, but this would mean measuring
rotation periods anywhere from $\sim$30 days to as long as $\sim$150--170 days. Such
extended rotation periods require rather lengthy observing campaigns. As light amplitudes
can be below $\sim$0.015 mag, the photometry requires a sufficiently high precision as
well. Further, successfully detecting a rotation signal depends on the star maintaining
a `favorable' (non-uniform) distribution of starspots, possibly for several years.
Fortunately, many of the M dwarfs observed in this program have displayed a persistence
of surface features -- in some cases for several years. Though most of our targets are
significantly older, our results thus far align with those of \citet{2020ApJ...897..125R}, which studied 4 young,
rapidly rotating M dwarfs.

\begin{figure}[ht!]
\plotone{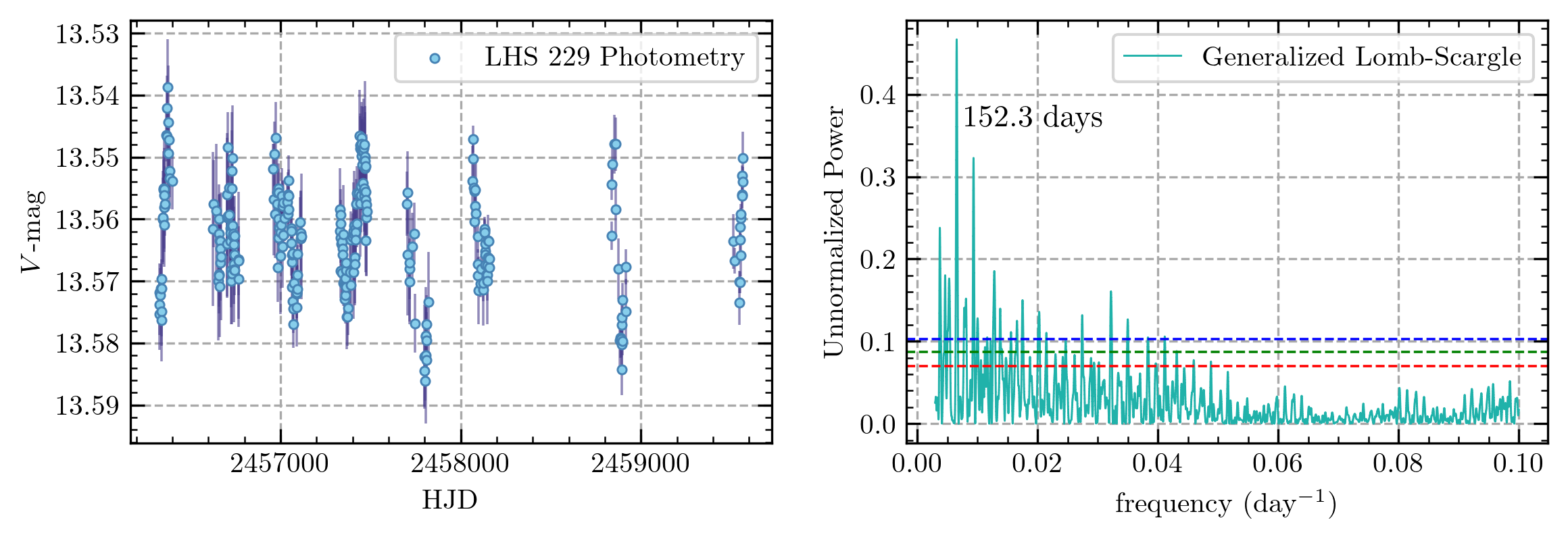}
\caption{An example lightcurve for one of our mid-late benchmark stars, LHS 229, whose age of 
$\sim$8.7 Gyr (see Table \ref{table:midmresults}) was determined from its WD companion, LHS 230
(see Table \ref{table:wdinfo}). At left, the full time-series dataset is plotted, covering a span
of 10 years. At right, the Generalized Lomb-Scargle periodogram (via \texttt{astropy}) results
are plotted, in frequency-space, and the dashed, horizontal lines indicate false alarm probabilities
(FAPs) of 10\% (red), 1\% (green), and 0.1\% (blue). A rotation period of $\sim$152.3 days was
found, and the phased lightcurve is plotted in Fig. \ref{fig:latemrots1}.
\label{fig:lhs229phot}}
\end{figure}

Apart from cluster members, whose rotation periods were obtained from the literature (see
Tables \ref{table:earlymresults} and \ref{table:midmresults}), the vast majority of benchmark
rotation periods were determined through dedicated CCD photometry of the targets carried out
with the 1.3 meter \textit{Robotically Controlled Telescope} \citep[\textit{RCT --}][]{2014AJ....147...49S} at 
\textit{Kitt Peak National Observatory} in Arizona. In limited cases, data (or additional data) were
obtained using other telescopes (e.g., the 0.8 meter \textit{Automated Photoelectric Telescope} (\textit{APT}) --
see \citealt{2015PhDT........45E}) or publicly available surveys, either due to target visibility or to help
confirm the rotation period. Except for the faintest sources, photometry was carried out in the \textit{V}-band,
with \textit{individual} measurement uncertainties of typically $\sim$0.004--0.01 mag depending on the
brightness of the star. Data would be removed prior to analysis for reasons ranging from legitimate hardware
malfunctions, to poor sky quality conditions, even down to a large moth having taken a poorly timed stroll
through the telescope's light path. For \textit{RCT} / \textit{APT} targets, 3 -- 5 measures were obtained
per night, from which nightly means and uncertainties were determined. The yearly and multi-year data sets were
searched for periodic variations with the Generalized Lomb-Scargle and CLEANest algorithms,
(as implemented within \texttt{astropy} \citep{2013AA...558A..33A,2018AJ....156..123A,2022ApJ...935..167A}
and \texttt{Peranso} (v3) \citep{2016AN....337..239P}) an example of which is shown in Fig. \ref{fig:lhs229phot}.
All rotation signals have false alarm probabilities below 1\%.

In limited cases, additional or alternative sources of photometry were used to determine
rotation periods. Proxima Cen and Kapteyn's Star have been a part of the program for several
years, but are southern hemisphere targets and thus inaccessible to the \textit{RCT}. Observing
time on the Skynet Robotic Telescope Network was purchased, and CCD photometry was carried out
using the network's \textit{PROMPT} telescopes at the \textit{Cerro Tololo Inter-American Observatory}
in Chile and the \textit{Meckering Observatory} in Australia. Images were automatically reduced before
downloading, and nightly means were determined and analyzed in similar fashion to \textit{RCT} / \textit{APT}
data.

The remaining cases consist of targets that were added to the program only recently. Due to the
installation and testing of a new and upgraded camera system, and the Contreras wildfire, the \textit{RCT}
experienced an observing hiatus. With a good combination of photometric depth, precision, and timeline,
\textit{Zwicky Transient Facility} data were used as an alternative. Individual measures were downloaded
from the \textit{NASA/IPAC} Data Archive, excluding any measures flagged by the archive or obtained at
high airmass. The resulting
data were then sigma-clipped (also within \texttt{astropy}) before nightly means were constructed. One example
is the open cluster NGC 752, where \textit{ZTF} photometry was used to both confirm the rotation periods of
\citet{2018ApJ...862...33A}, and provide additional periods (see Table \ref{table:ngcruprecht}. Similarly, we were able to measure rotation
periods for a small sample of Ruprecht 147 mid-late M dwarfs using \textit{ZTF} photometry. Members of
both clusters are now being observed with the \textit{RCT} to further confirm previous results, but more
specifically to expand the inventory of reliably determined rotation periods.

\begin{figure}[ht!]
\plotone{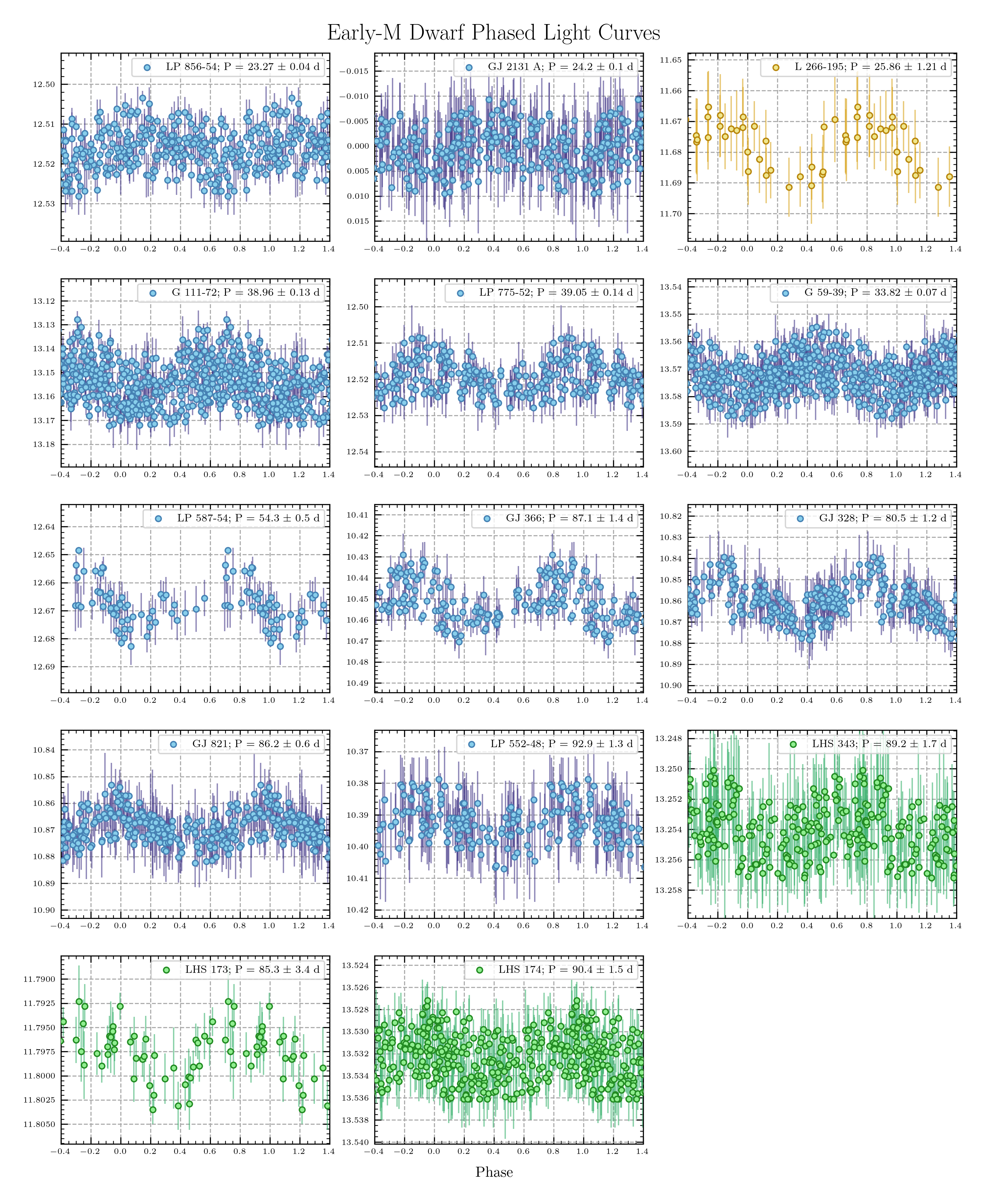}
\caption{The phased rotation light curves of the early M dwarf benchmarks. The name and rotation period
(and uncertainty) of each star is inset within their respective plot. The `scatter' of the data
along the y-axis is primarily due to variations in lightcurve shape and amplitude over the different
observing seasons. For plots where the magnitudes are centered on zero, either a long-term trend or
cycle was first removed from the data prior to conducting the rotation period search. Data are color-coded according
to their source: blue for data that we acquired with the \textit{RCT} (and \textit{APT} in limited
cases), green for public \textit{ZTF} data \citep{https://doi.org/10.26131/irsa538}, and beige for data we obtained with the \textit{Skynet}
telescope network. \label{fig:earlymrots}}
\end{figure}

Rotational light curves of the benchmark M dwarfs are shown in Fig. \ref{fig:earlymrots} to
give an idea of the amplitudes of variability observed and of the data quality. M dwarf
rotation amplitudes can range from $\sim$0.01 mag or less in the ``toughest'' cases to as much as
0.06 mag for ideal cases such as Proxima Cen.

Several additional potential benchmarks have been observed, and continue to be, but are not included in
this study. Some simply have yet to show a coherent rotation period. For others (e.g., LHS 3279), more
than one potential rotation period is observed and further data will hopefully reveal true rotation
period. There are others still for which rotation periods have been found that are at odds with the
currently determined age, but perhaps with reason. G 121-21 is one example, where it has an age
(determined via its white dwarf companion) of $\sim$2 Gyr, yet \textit{RCT} photometry revealed a rotation
period of $\sim$0.62 days which we confirmed by analyzing \textit{TESS} photometry. Though it's
possible this star represents an extreme example of the spread of rotation periods at younger ages,
we believe that an unresolved companion orbiting the M dwarf is more likely the cause of its rapid rotation.

\section{Results \& Discussion -- A Tale of Two Relationships} \label{sec:highlight}

\begin{figure}[ht!]
\plotone{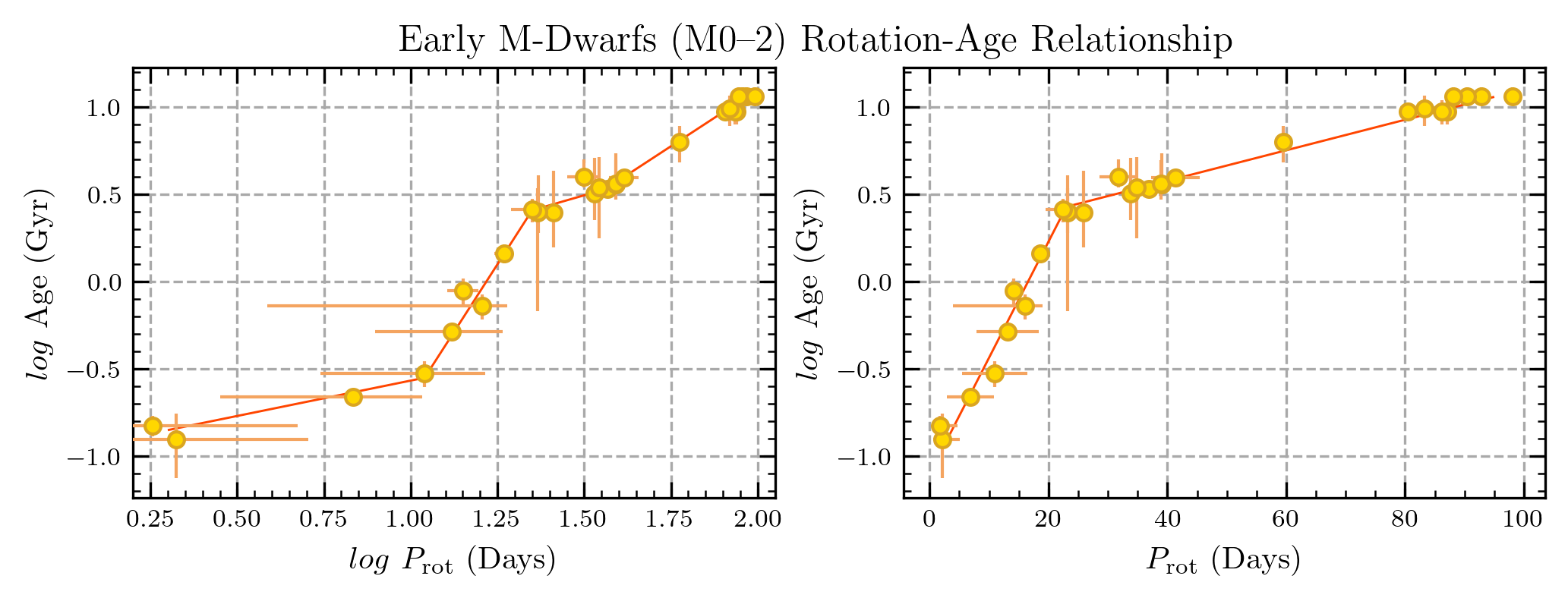}
\plotone{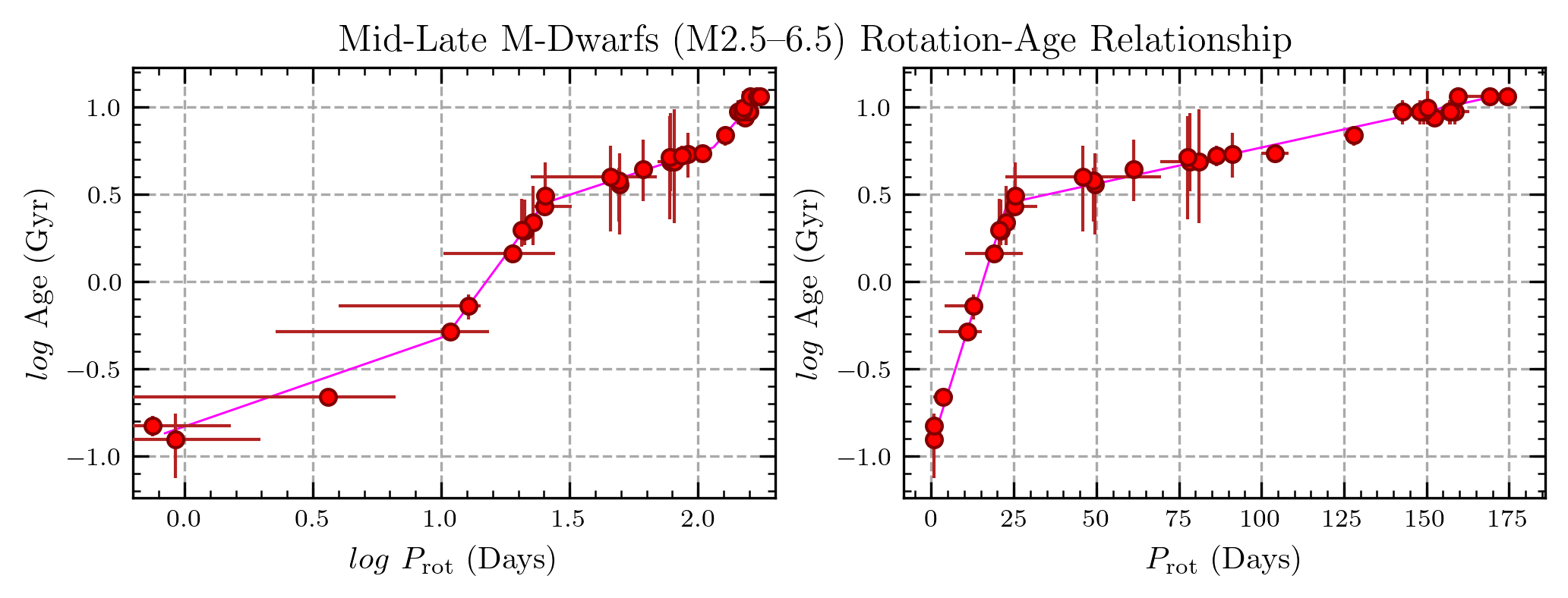}
\caption{Plots showing log-log (left) vs semi-log (right) Age vs. Rotation
relationships for the early (top) and mid-late (bottom) M dwarfs. As shown in the plots, the data of
both groups are better linearized in semi-log form. It is perhaps possible that each `segment' of the
log-log plots represents a real evolutionary stage -- e.g., the first segment is pre-main sequence
evolution and the second represents the interiors of the M dwarfs synchronizing -- and the semi-log plots
merge these two distinct stages. However, given the current data and spread of rotation rates in the
clusters, such a firm conclusion can't be drawn here. \label{fig:rotagelinearize}}
\end{figure}

The primary focus of the \textit{Living with a Red Dwarf}
(\textit{LivRed}) program has been to characterize the evolution of M dwarf rotation rates 
over their lifetimes, with the end goal of providing a reliable method for calculating the age
of an M dwarf, so long as its rotation period has been determined. When comparing related, age-associated
quantities (e.g., activity and rotation) the data is commonly linearized by taking the logarithms of
both quantities. In our analysis of M dwarf age vs rotation, we found that both subsets of M dwarfs 
(but particularly the mid-late subset) showed deviations from linearity in log-log space, and a 
more straightforward analysis of their rotations over time could be carried out in semi-log space 
(see Fig \ref{fig:rotagelinearize}) while clearly showing the inflection points on the evolutionary
tracks of both early and mid-late M dwarfs that will be discussed later in this section. 

As mentioned previously, constructing these relationships
proved a rather complicated task, but not simply due to the difficulty in building a substantial set
of benchmark targets and the observational burden of measuring their rotation periods. When it comes
to spindown, it appears there is no way to broadly classify all M dwarfs; they represent too wide a
range of parameters. 

If you group together G dwarfs, the masses can vary by $\sim$10\%. Grouping together all K dwarfs,
the mass can vary by $\sim$30 -- 40\%. But studying M dwarfs, even focusing only on M0 -- 6.5 dwarfs as we
have done here, the mass can vary by $>500\%$. Important changes occur within the stars' interior
structures and evolutionary timelines, which can be observed (and need to be accounted for) in the
rotation relationships.

One dramatic difference was suspected early in the study, but required lengthy follow-up. Preliminary
project results showed the rotations of (especially older) early vs. mid-late M dwarfs followed divergent
evolutionary paths \citep{2018RNAAS...2...34E} as they aged. This splitting of M dwarf subsets into
distinct rotation-based groups has been observed in other studies, as well (see \citealt{2021ApJ...916...77P}).
Also particularly relevant to this study; models have shown (Louis Amard, private communication) that
the oldest stars (subdwarf members of the Halo population) display an interesting and related phenomenon
where their interiors are structured as that of a main sequence star with slightly later spectral type
(see Tables \ref{table:earlymresults} and \ref{table:midmresults}. An example is Kapteyn's Star, which
would initially be considered a member of our early subset since it is classified as sdM1.5, yet
models indicate it has a fully convective interior similar to a $\sim$M2.5 or later main sequence star.
A potential explanation for subdwarfs having deeper convective zones than their spectral types would
indicate is that their smaller radii lead to larger interior temperature gradients, but confirming the
true cause requires further study.

\begin{figure}[ht!]
\plotone{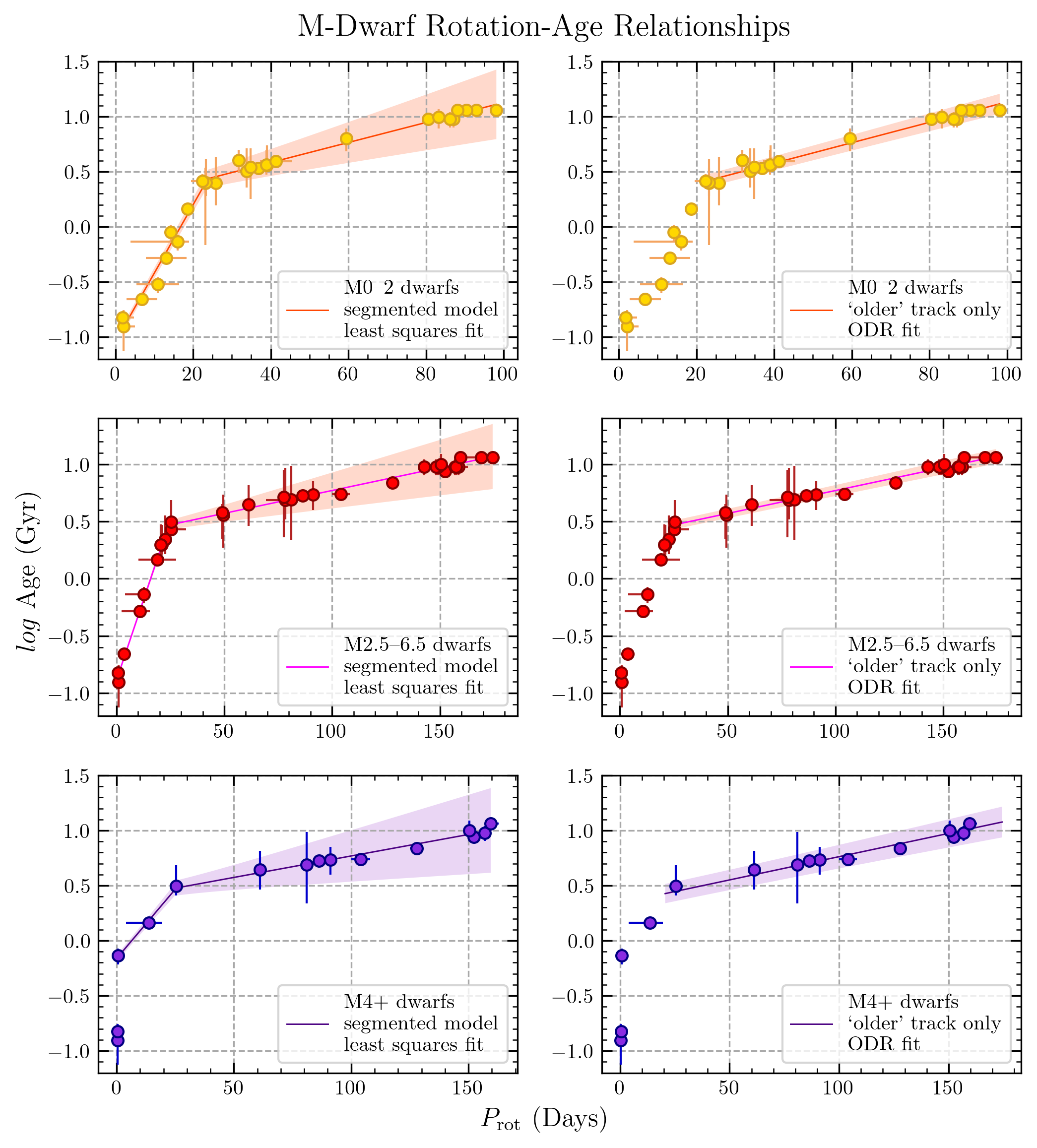}
\caption{Age-Rotation relationships are plotted for the different M dwarf subsets, in semi-log space. The
left-hand plot shows the segmented linear fit for each stellar subset's full time span of data, while the
right-hand plot shows the single linear fit of only the subset's older evolutionary track. In the
top panels, only the early (M0 -- 2) subset is plotted. In the middle panels, the mid-late subset is plotted. This subset
contains M2.5 -- 6.5 dwarfs except on the young track where only M2.5 -- 3.5 dwarfs are plotted. In the bottom
panels, only the data for M4 and later spectral types are plotted.
Clear inflection points can be seen in the plots of all stellar subsets. This is believed to occur when the interiors
of the stars re-synchronize, and the next phase of their rotational evolution can begin. Also note the
drastically different rotation-scales for each subset, with the early M dwarfs slowing to ~100 days,
yet the mid-late M dwarfs slow much further, to ~175 day rotation periods. \label{fig:rotagesemilog}}
\end{figure}

By an age of 10 Gyr, the average mid-late M dwarf will have a rotation period almost twice as long as
the average early M dwarf ($\sim$155 vs $\sim$85 days -- see Figs \ref{fig:rotagelinearize} \& \ref{fig:rotagesemilog}).
These different paths resulted in our first subdivision of the M dwarfs, into what we call the `early'
(M0 -- 2) and `mid-late' (M2.5 -- 6.5) groups. This is near to, but earlier than, the usual spectral type of
M3 -- 3.5 which is routinely quoted as the transition point to a fully convective interior. \citet{2020ApJ...891..128M}
recently showed, however, that changes due to magnetic effects (which would be rotation-related) likely
occur within the M2.1 -- 2.3 range, which is encouraging in light of our rotation period results.
The decision to not include stars with spectral types later than $\sim$M6.5 V is intentional. First,
we presently do not have a sufficient sample of older stars at these later spectral types with both
well-determined ages and rotation periods. Second, from the small sample of such stars that is
available, it appears they either experience no appreciable spindown effect, or one that is altogether
different from any of the other M dwarf subsets presented here. A well-known example is the M8 V star
Trappist-1, that has an age of 7.6 $\pm$ 2.2 Gyr and a rotation period of 3.39 days
(e.g., \citealt{2017ApJ...845..110B}). 

There is an additional complication at young ages, due to the length of time required for each
star to reach the main sequence. As with other issues, this one is also particularly difficult 
for M dwarfs, whose pre-main sequence lifetimes can
range from $\sim$140 Myr to $\sim$1.5 Gyr (M0 – 6.5 dwarfs \citealt{2016ApJ...823..102C}). Due to this,
on top of the spread of initial rotation rates normally displayed by all cool dwarf stars,
younger M dwarfs show a larger range of rotation periods, as detailed in several excellent
studies (see \citealt{2019ApJ...879..100D,2020AA...644A..16G,2020ApJ...904..140C,2021ApJS..257...46G}
for recent examples). After an age of $\sim$2.5 -- 3 Gyr, all mid-late M dwarfs have converged onto a single
evolutionary track, and any differences between their rotation-determined ages are negligible compared
to the uncertainties of the relationships. At young ages, though, this group appears to require further
subdivision, most likely as a result of lengthening pre-main sequence lifetimes. Due to the distances of
some of these young clusters, this aspect of the study is in need of further study, but M dwarfs later
than $\sim$M4 sensibly appear to follow a different evolutionary path while young (the middle and bottom plots of
Fig \ref{fig:rotagesemilog} show this further subdivision).

Previous studies (see \citealt{2020ApJ...904..140C}) have proposed the rotation periods of cool
dwarfs do not follow one continuous evolutionary track, characterized by a single powerlaw
relationship ($P_{\rm{rot}}$ $\propto$ $Age^n$), with a `braking index' determined by the
exponent (\textit{n}). This was originally proposed by \citet{1972ApJ...171..565S}, with an
initially determined value of \textit{n} = 0.5 for that study's target sample, and recent
studies have revised this value. \citet{2019ApJ...879..100D} and \citet{2022ApJ...938..118D},
for example, each derived an index of $n \approx$ 0.62,
although \citeauthor{2019ApJ...879..100D} studied only F and G dwarfs -- more massive than the
stars studied here -- but \citeauthor{2022ApJ...938..118D} included stars up to $\sim$M3.

For these data, and the activity-age relationships in the following subsections, two fits were
applied. A two-segment linear equation was defined using the 
\texttt{numpy.piecewise} function and fit to the full age-range using \texttt{scipy.optimize.least\_squares},
and an unsegmented linear model was fit to only the `older' track of each dataset via orthogonal distance
regression (using the \texttt{BCES}\footnote{\url{https://github.com/rsnemmen/BCES}} module written
by Rodrigo Nemmen). With fewer terms, and fitting only the timespan after all mid-late M dwarfs appear
to have converged onto a single evolutionary path, the unsegmented model shows significantly reduced
uncertainties (as seen in the following equations, and in the ages calculated in Table \ref{table:host_ages}).
The final, fitted age-rotation relationships are shown in Fig \ref{fig:rotagesemilog} with the best-fitting parameters:


\vspace{5mm}

M0--2 dwarfs -- segmented fit of full age range:
\begin{align}
\log Age\; (Gyr) = &0.0621 [0.0024] \times P_{\rm{rot}}\; (days) - 1.0437 [0.0380] \notag\\ 
&\textrm{for} ~P_{\rm{rot}} < 23.4933 [0.7643] \notag\\
\log Age\; (Gyr) = &0.0621 [0.0024] \times P_{\rm{rot}}\; (days) - 1.0437 [0.0380] \notag\\ 
- &0.0528 [0.0025] \times (P_{\rm{rot}} - 23.4933 [0.7643]) \notag\\ 
&\textrm{for} ~P_{\rm{rot}} \geq 23.4933 [0.7643]
\end{align}

M0--2 dwarfs -- linear ODR fit to the `older track' only:
\begin{align}
\log Age\; (Gyr) = &0.0094 [0.0002] \times P_{\rm{rot}}\; (days) + 0.1909 [0.0114] \notag\\
&\textrm{for} ~P_{\rm{rot}} \gtrsim 22
\end{align}

M2.5--6.5 dwarfs  [M2.5 -- M3.5 for P$_{\rm{rot}} \lesssim$ 24]: 
\begin{align}
\log Age\; (Gyr) = &0.0561 [0.0012] \times P_{\rm{rot}}\; (days) - 0.8900 [0.0185] \notag\\
&\textrm{for} ~P_{\rm{rot}} < 24.1888 [0.4268] \notag\\
\log Age\; (Gyr) = &0.0561 [0.0012] \times P_{\rm{rot}}\; (days) - 0.8900 [0.0185] \notag\\
- &0.0521 [0.0012] \times (P_{\rm{rot}} - 24.1888 [0.4268]) \notag\\
&\textrm{for} ~P_{\rm{rot}} \geq 24.1888 [0.4268]
\end{align}

M2.5--6.5 dwarfs -- linear ODR fit to the `older track' only:
\begin{align}
\log Age\; (Gyr) = &0.0041 [0.0001] \times P_{\rm{rot}}\; (days) + 0.3691 [0.0126] \notag\\
&\textrm{for} ~P_{\rm{rot}} \gtrsim 22
\end{align}

M4 -- 6.5 dwarfs:
\begin{align}
\log Age\; (Gyr) = &0.0251 [0.0018] \times P_{\rm{rot}}\; (days) - 0.1615 [0.0303] \notag\\
&\textrm{for} ~P_{\rm{rot}} < 25.4500 [1.9079] \notag\\
\log Age\; (Gyr) = &0.0251 [0.0018] \times P_{\rm{rot}}\; (days) - 0.1615 [0.0303] \notag\\
- &0.0212 [0.0018] \times (P_{\rm{rot}} - 25.4500 [1.9079]) \notag\\
&\textrm{for} ~P_{\rm{rot}} \geq 25.4500 [1.9079]
\end{align}

M4--6.5 dwarfs -- linear ODR fit to the `older track' only:
\begin{align}
\log Age\; (Gyr) = &0.0042 [0.0002] \times P_{\rm{rot}}\; (days) + 0.3401 [0.0288] \notag\\
&\textrm{for} ~P_{\rm{rot}} \gtrsim 25
\end{align}

Together, these relationships cover M0--6.5 dwarfs. Within 10 pc of the Sun, this represents $\sim$72\% of all
stars with known spectral types (and $\sim$48\% of the wider range of objects, including
stellar remnants and brown dwarfs \citet{2021AA...650A.201R}). As mentioned early in the paper,
an increasing number of M dwarfs are being discovered as exoplanet hosts. Determining the ages
of these stars, and thus the ages of their exoplanets, is important when selecting the ideal
targets to further study for evidence of habitability or even life. Single-celled organisms
originated when the Earth (the only example we currently have for such events) was $\sim$0.7 -- 0.9
Gyr old. The Great Oxygenation of the atmosphere occurred when Earth was $\sim$2.2 Gyr old,
the Cambrian Explosion and rapid diversification of complex lifeforms when Earth was $\sim$4 -- 4.1 Gyr old,
and technological civilization didn't occur until the Earth was $>$4.5 Gyr old. Thus, exoplanet
age is an important descriminator in the search for life. To demonstrate a benefit
of our relationships, we provide the gyrochronological ages for all $\sim$M0--6.5
dwarf exoplanet hosts with a listed rotation period in the 
\textit{NASA Exoplanet Archive\footnote{\url{https://exoplanetarchive.ipac.caltech.edu/}}} in
Table \ref{table:host_ages}. The follow-up paper (Engle 2023) will focus on the X-ray and UV
activity of M dwarfs over time, and what insights these relationships offer for the stars, their
magnetic dynamos, and their suitability to host potentially habitable planets.

We do wish to advise restraint, though, when using the first or `young' tracks of the segmented relationships.
These time spans were included to, as best we could, characterize the fullest evolutionary paths of average
early to mid-late M dwarfs. However, as noted earlier in this section, M dwarfs display a \textit{wide} range
of rotation rates at young ages that are not represented by the relationship uncertainties (note the data point
 vs. relationship uncertainties along the `young' tracks in Fig \ref{fig:rotagesemilog}). Fortunately, estimates
 from \textit{activity}-age relationships can extend the reliable range of age determinations into this younger
 time span, and this will be shown and discussed in the companion paper. Finally, while the
 segmented relationships determine young age uncertainties that are likely too small, including this time span 
 ironically inflates the uncertainties of the older age range. For this time span, additional age and uncertainty
 measures using fits to only the older data will be informative.

\citet{2003ApJ...586..464B} theorized that the rotational evolution of main sequence stars showed
two possible sequences, dependent on both time and mass (therefore spectral
type). The two sequences were termed the Interface (I) and Convective (C) sequences,
named after the magnetic dynamos and interior structures of the stellar groups. The more
massive, hotter stars spend only a short time on the C sequence before switching to the
I sequence. On the I sequence, there is an interface between the radiative and convective
regions of the stellar interior, but the magnetic field couples the regions together and
much of the stellar interior rotates as a rigid body. Lower mass stars spend longer
amounts of time on the C sequence before switching over, and fully convective stars
likely never leave the C sequence. \citet{2020AA...636A..76S} put forth an analytical model
theorizing that stars do not initially evolve as rigid bodies. Rather, as the stellar surface
loses angular momentum, a profile of differential rotation builds within the star. Angular 
momentum is transported from the interior to the surface and eventually the interior of the star
`re-couples'. This accounts for the two-track evolutionary path where the second evolutionary
track begins after the interior of the star has re-coupled. \citeauthor{2020AA...636A..76S}
only calculated their model down to early M dwarf masses. At this mass range, however, the
models predict that a $\sim$4 Gyr old early M dwarf will have a $\sim 31-32$ day
rotation period, where our data indicates a $\sim40-45$ day period.

To determine the braking indices of our M dwarf subsets, and to serve as an additional
comparison to literature results, rotation vs. age data were fitted in linear space
with a two-segment powerlaw equation. A best fitting braking index (for the second
evolutionary track) was determined to be 0.61 for the early M dwarfs and 0.62 for the
mid-late M dwarfs; nearly identical to each other, well within the parameter uncertainties,
and in excellent agreement with the results of \citet{2019ApJ...879..100D} and \citet{2022ApJ...938..118D}. However, it is
again worth noting that \citealt{2019ApJ...879..100D} based their braking index determination
on solar-like F and G dwarfs, and on a comparison of the Praesepe cluster and the Sun. Just 
over fifty years after \citet{1972ApJ...171..565S} first discovered the spindown effect
operating in solar-type G dwarfs, it is an interesting
implication that all cool dwarfs, from late F to $\sim$M6.5, may perhaps spindown according
to the same braking index but simply with different re-coupling timescales. Further 
investigation into the angular momentum loss of M dwarfs, using methods such as those
of \citet{2019ApJ...886..120S} and \citet{2010ApJ...721..675B} and comparisons between
measures and estimated magnetic field strengths and mass loss rates, are underway for
inclusion in a follow-up paper.

These data and relationships will be of great use to the field and offer
valuable insights into the most populous stellar members of our galaxy, M dwarfs. They
allow for reliable ages and evolutionary histories to be determined, but may also offer further
insight into the differing dynamo mechanisms at work within the M dwarf subsets and how
each mechanism influences, or responds to, the star's evolution over time.

\begin{figure}[ht!]
\plotone{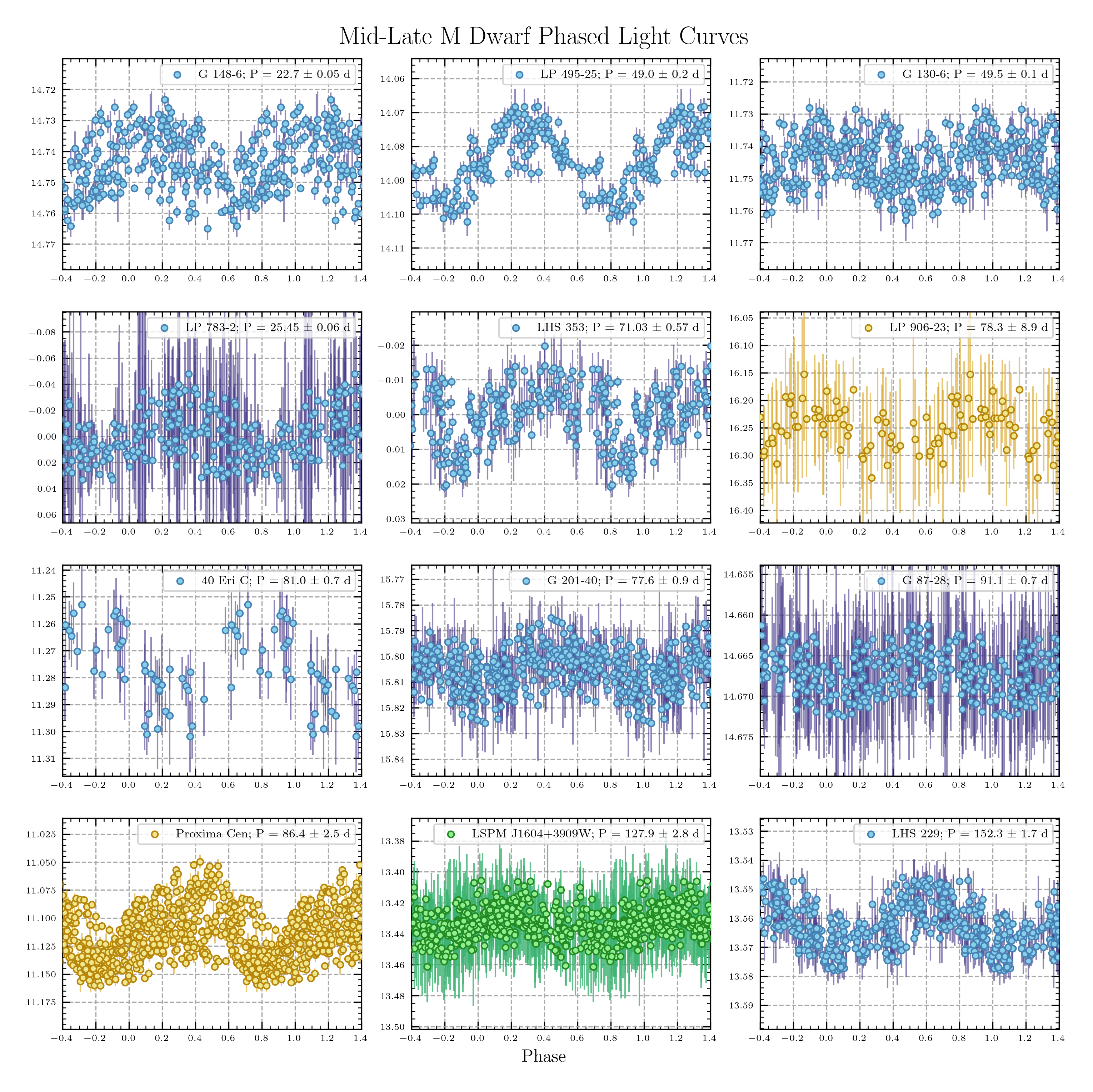}
\caption{The phased rotation light curves of the mid-late M dwarf benchmarks (additional lightcurves are
plotted in Fig. \ref{fig:latemrots2}). The data coloring scheme from Fig. \ref{fig:earlymrots} is applied
here and, again as in Fig. \ref{fig:earlymrots}, a plot where magnitudes are centered on zero indicates
that a long-term trend or cycle has been removed.
\label{fig:latemrots1}}
\end{figure}

\begin{figure}[ht!]
\plotone{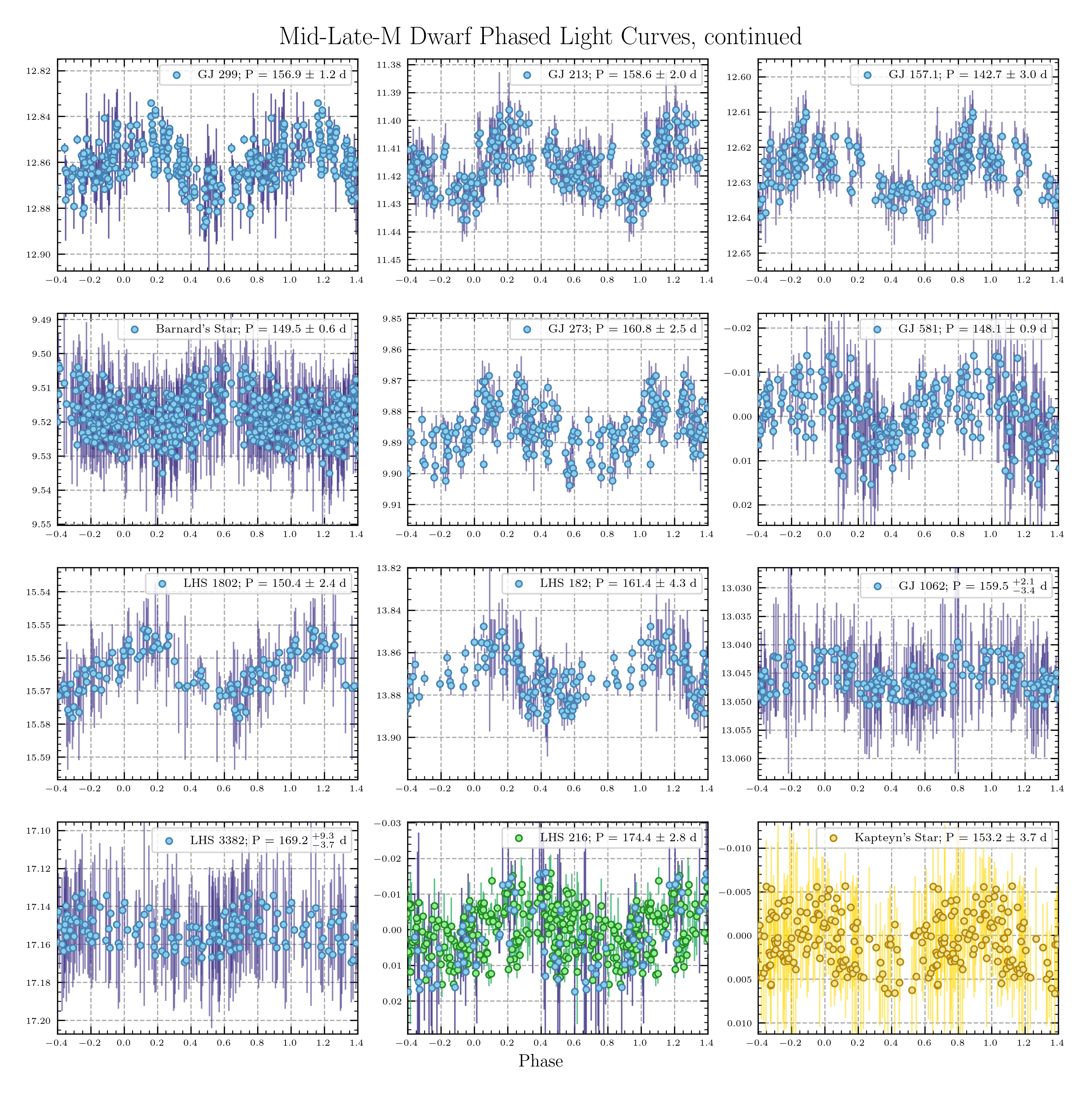}
\caption{Additional phased rotation light curves of the mid-late M dwarf benchmarks, continued from
and plotted in a similar fashion as Fig. \ref{fig:earlymrots}.  \label{fig:latemrots2}}
\end{figure}

\begin{deluxetable*}{ll}
\tablecaption{Rotations Periods in Addition to Those Found in the Literature for NGC 752 and Ruprecht 147 Members \label{table:ngcruprecht}}
\tablehead{\colhead{Star Name}  & \colhead{$P_{\rm{rot}}$ (days)}}
\startdata
\multicolumn{2}{c}{NGC 752}           \\
\hline
2MASS   J01561789+3757110      & 11.0 \\
2MASS   J01572038+3747404      & 18.2 \\
2MASS   J01583743+3750269      & 12.6 \\
2MASS   J01584750+3734036      & 10.6 \\
2MASS   J01570853+3750213      & 50.9 \\
2MASS   J01572291+3757466      & 8.7  \\
2MASS   J01570025+3747243      & 25.0 \\
2MASS   J01582469+3733380      & 21.7 \\
\hline
\multicolumn{2}{c}{Ruprecht 147}      \\
\hline
Gaia   DR2 4179511879236506112 & 27.1 \\
Gaia   DR2 4199836897135629440 & 28.4 \\
Gaia   DR2 4180583692619791360 & 20.0 \\
Gaia   DR2 4184987580281615616 & 24.9 \\
Gaia   DR2 4184292792011931776 & 32.6 \\
\enddata
\tablecomments{All rotation periods have uncertainties of 0.1 days or below, but we are reporting the values to this
level of precision until the periods are confirmed through dedicated, follow-up photometry.}
\end{deluxetable*}

\begin{deluxetable*}{lllll}
\tablecaption{The `Early' M Dwarf (M0--2) Benchmarks \label{table:earlymresults}}
\tablehead{\colhead{Name}  & \colhead{Sp. Type}  & \colhead{Age (Gyr)}  & \colhead{$P_{\rm{rot}}$ (days)}  & \colhead{Age determined   via}}
\startdata
Pleiades                  & M0--2V    & 0.125 {[}$+0.05, -0.05${]}   & 2.11 {[}$+2.96,   -1.50${]}$^1$  & Cluster                      \\
NGC 2516                  & M0--2V    & 0.15 {[}$+0.02, -0.02${]}    & 1.80 {[}$+2.93,   -1.30${]}$^1$  & Cluster                      \\
M34                       & M0--2V    & 0.22 {[}$+0.02, -0.02${]}    & 6.82 {[}$+3.99,   -3.99${]}$^1$  & Cluster                      \\
NGC 3532                  & M0--2V    & 0.30 {[}$+0.05,  -0.05${]}   & 10.96 {[}$+5.47,   -5.47${]}$^2$ & Cluster                      \\
M37                       & M0--2V    & 0.52 {[}$+0.06, -0.06${]}    & 13.15 {[}$+5.24, -5.24${]}$^1$   & Cluster                      \\
Praesepe / Hyades         & M0--2V    & 0.73 {[}$+0.12, -0.12${]}    & 16.06 {[}$+2.91, -12.2${]}$^3$   & Cluster                      \\
NGC 6811                  & M0--2V    & 0.89 {[}$+0.15, -0.15${]}    & 14.19 {[}$+1.45, -1.45${]}$^4$   & Cluster                      \\
NGC 752                   & M0--2V    & 1.46 {[}$+0.18, -0.18${]}    & 18.6 {[}$+4.2, -4.2${]}$^5$      & Cluster                      \\
L 266-195                 & M1V       & 2.50 {[}$+1.82,   -0.93${]}  & 23.19 {[}$+1.84,   -1.84${]}     & WD comp (L 266-196)          \\
LP 856-54                 & M1--1.5V  & 2.52 {[}$+1.58,   -0.61${]}  & 23.27 {[}$+0.04,   -0.04${]}     & WD comp (LP   856-53)        \\
Ruprecht   147            & M0--2V    & 2.6 {[}$+0.4, -0.4${]}       & 22.4 {[}$2.9, -2.9${]}$^{6,7}$   & Cluster                      \\
G 59-39                   & M0V       & 3.20 {[}$+1.96, -0.93${]}    & 33.82 {[}$+0.07,   -0.07${]}     & WD comp (EGGR 92)            \\
LP 775-52                 & M0--1V    & 3.39 {[}$+0.29, -0.20${]}    & 36.90 {[}$+0.10,   -0.10${]}     & WD comp (LP   775-53)        \\
GJ 2131 A                 & M1V       & 3.47 {[}$+1.69,   -1.69${]}  & 34.82 {[}$+0.10, -0.10${]}       & WD comp (GJ 2131   B)        \\
G 111-72                  & M1.5--2V  & 3.64 {[}$+1.35,   -0.68${]}  & 38.96 {[}$+0.13,   -0.13${]}     & WD comp (G   111-71)         \\
HIP 43232 B               & M1.5V     & 3.95 {[}$+0.35,   -0.35${]}$^{8}$    & 41.3 {[}$+4.1, -4.1${]}$^{8}$  & MS comp (HIP   43232 A)   \\
M67                       & M0--2V    & 4.0 {[}$+1.0, -0.5${]}       & 31.8 {[}$+3.3, -3.3${]}$^{9}$          & Cluster                      \\
LP 587-54                 & M1.5--2V  & 6.34 {[}$+1.49,   -1.38${]}  & 59.5 {[}$+1.4, -1.4${]}          & WD comp (LP   587-53)        \\
GJ 366                    & M1.5V     & 9.5 {[}$+1.5, -1.5${]}       & 87.1 {[}$+1.2, -1.2${]}          & Thick Disk/Halo   Population \\
GJ 328                    & M0V       & 9.5 {[}$+1.5, -1.5${]}       & 80.5 {[}$+0.6, -0.6${]}          & Thick Disk/Halo   Population \\
GJ 821                    & M1V       & 9.5 {[}$+1.5, -1.5${]}       & 86.2 {[}$+1.1, -1.1${]}          & Thick Disk/Halo   Population \\
LP 552-48                 & M0V       & 9.88 {[}$+1.78, -2.06${]}    & 83.2 {[}$+0.8, -0.8${]}          & WD comp (LP   552-49)        \\
LHS 343                   & sdK7      & 11.5 {[}$+1.0, -1.5${]}      & 89.2 {[}$+1.7, -1.7${]}          & Halo Population              \\
LHS 173                   & esdK7$^{10}$ & 11.5 {[}$+1.0, -1.5${]}     & 85.3 {[}$+3.4, -3.4${]}          & Halo Population               \\
LHS 174                   & sdM0$^{10}$ & 11.5 {[}$+1.0, -1.5${]}      & 90.4 {[}$+1.5, -1.5${]}          & Halo Population               \\
\enddata
\tablecomments{\footnotesize{$^1$\citet{2021ApJS..257...46G}   $^2$\citet{2021AA...652A..60F}   $^3$\citet{2022ApJ...931...45N}  $^4$\citet{2019ApJ...879...49C}  $^5$\citet{2018ApJ...862...33A}, with additional rotation periods measured by us using \textit{ZTF} photometry  $^{6,7}$\citet{2020ApJ...904..140C,2020AA...644A..16G}  $^{8}$Sawczynec, E. (2021), Thesis available at \url{https://www.phys.hawaii.edu/wp-content/uploads/2021/06/ESawczynec_Thesis-1.pdf}    $^{9}$\citet{2022ApJ...938..118D}    $^{10}$\citet{2019AJ....157...63K}}}
\end{deluxetable*}

\begin{longrotatetable}
\begin{deluxetable*}{lllll}
\tablecaption{The `Mid' M Dwarf (M2.5--6.5) Benchmarks \label{table:midmresults}}
\tablehead{\colhead{Name}  & \colhead{Sp. Type}  & \colhead{Age (Gyr)}  & \colhead{$P_{\rm{rot}}$ (days)}  & \colhead{Age determined via}}
\startdata
Pleiades                  & M2.5--$\sim$3.5V$^1$   & 0.125 {[}$+0.05,   -0.05${]}& 0.92 {[}$+1.06,   -0.43${]}$^2$ & Cluster        \\
Pleiades                  & M4--$\sim$6.5V$^1$     & 0.125 {[}$+0.05,   -0.05${]}& 0.38 {[}$+0.15,   -0.10${]}$^2$ & Cluster        \\
NGC 2516                  & M2.5--$\sim$3.5V       & 0.15 {[}$+0.02,   -0.02${]} & 0.68 {[}$+0.77,   -0.30${]}$^2$ & Cluster        \\
NGC 2516                  & M4--6V                 & 0.15 {[}$+0.02,   -0.02${]} & 0.41 {[}$+0.09,   -0.10${]}$^2$ & Cluster              \\
M34                       & M2.5--$\sim$3.5V       & 0.22 {[}$+0.02,   -0.02${]} & 3.63 {[}$+3.01,   -3.01${]}$^2$ & Cluster              \\
M37                       & M2.5--6V               & 0.52 {[}$+0.06,   -0.06${]} & 10.84 {[}$+4.55,   -8.57${]}$^2$ & Cluster             \\
Praesepe / Hyades         & M2.5--$\sim$3.5V$^1$   & 0.73 {[}$+0.12,   -0.12${]} & 12.81 {[}$+1.42, -8.83${]}$^3$ & Cluster         \\
Praesepe / Hyades         & M4--$\sim$6.5V$^1$     & 0.73 {[}$+0.12,   -0.12${]} & 23.8 {[}$+3.3, -3.3${]}$^3$ & Cluster          \\
NGC 752                   & M2.5--$\sim$3.5V       & 1.46 {[}$+0.18,   -0.18${]} & 19.83 {[}$+10.93,   -10.93${]}$^4$ & Cluster           \\
NGC 752                   & M4--6V    & 1.46 {[}$+0.18,   -0.18${]}   & 13.7 {[}$+5.8,   -9.8${]}$^4$ & Cluster                      \\
NSV 11919                 & M2.5--3V  & 1.97 {[}$+0.98,   -0.35${]}   & 21.07 {[}$+0.05, -0.05${]}     & WD comp (NSV 11920)          \\
Gaia DR3 7552$^{5a}$      & M3--3.5V  & 1.98 {[}$+1.01, -0.39${]}     & 20.6 {[}$+1.1, -1.1${]}        & WD comp (Gaia DR3 4784$^{5b}$)  \\
G 148-6                   & M3--3.5V  & 2.20 {[}$+1.35, -0.57${]}     & 22.7 {[}$+0.05, -0.05${]}     & WD comp (G   148-7)          \\
Ruprecht   147            & M2.5--$\sim$4V  & 2.6  {[}$+0.4, -0.4${]}     & 26.6 {[}$+4.2, -4.2${]}$^{6}$     & Cluster                      \\
LP 783-2                  & M6.5V     & 3.12 {[}$+1.73, -0.56${]}   & 25.45 {[}$+0.06, -0.06${]}     & WD comp (LP 783-3)        \\
G 130-6                   & M3V       & 3.60 {[}$+1.84, -1.73${]}   & 49.5 {[}$+0.1, -0.1${]}     & WD comp (G   130-5)          \\
LP 498-25                 & M2.5V     & 3.80 {[}$+0.70, -1.58${]}   & 49.0 {[}$+0.2, -0.2${]}     & WD comp (LP   498-26)        \\
M67                       & M2--$\sim$3.5V    & 4.0 {[}$+1.0, -0.5${]}    & 46.0 {[}$+23.5, -23.5${]}$^{7}$   & Cluster          \\
LHS 353                   & M4V       & 4.42 {[}$+2.13, -1.52${]}   & 61.2 {[}$+2.1, -2.1${]}     & WD comp (GJ 515)             \\
LP 906-23                 & M4V       & 4.86 {[}$+4.41, -1.57${]}   & 78.3 {[}$+8.9, -8.9${]}     & WD comp (LP 906-24)          \\
40 Eri C                  & M4.5V     & 4.89 {[}$+4.81, -2.71${]}   & 81.0 {[}$+0.7, -0.7${]}     & WD comp (40 Eri   B)         \\
G 201-40                  & M3--3.5V  & 5.16 {[}$+3.77, -2.87${]}   & 77.6 {[}$+0.9, -0.9${]}     & WD comp (G   201-39)         \\
Proxima Cen               & M5.5V     & 5.3 {[}$+0.7, -0.7${]}      & 86.4 {[}$+2.5, -2.5${]}     & $\alpha$ Cen system          \\
G 87-28                   & M4V       & 5.42 {[}$+1.70, -1.45${]}   & 91.1 {[}$+0.7, -0.7${]}     & WD comp (G   87-29)          \\
2MASS J23095781+5506472   & M5.5--6V  & 5.47 {[}$+0.61, -0.26${]}   & 104.1 {[}--, --{]}          & WD comp (2MASS J23095848+5506491)   \\
LSPM J1604+3909W          & M4V       & 6.9 {[}$+0.9, -0.9${]}      & 127.9 {[}$+2.8,   -2.8${]}  & MS comp (HD   144579)        \\
LHS 229                   & M4V       & 8.71 {[}$+0.56, -0.46${]}   & 152.3{[}$+1.7, -1.7${]}     & WD comp (LHS   230)          \\
GJ 299                    & M4.5V     & 9.5 {[}$+1.5, -1.5${]}      & 156.9 {[}$+1.2, -1.2${]}    & Thick Disk/Halo   Population \\
GJ 213                    & M4V       & 9.5 {[}$+1.5, -1.5${]}      & 158.6 {[}$+2.0, -2.0${]}    & Thick Disk/Halo   Population \\
GJ 157.1                  & M4V       & 9.5 {[}$+1.5, -1.5${]}      & 142.7 {[}$+3.0, -3.0${]}    & Thick Disk/Halo   Population \\
Barnard's Star   (GJ 699) & M4V       & 9.5 {[}$+1.5, -1.5${]}      & 149.1 {[}$+0.6, -0.6${]}    & Thick Disk/Halo   Population \\
GJ 273                    & M3.5V     & 9.5 {[}$+1.5, -1.5${]}      & 157.3 {[}$+5.7, -5.7${]}    & Thick Disk/Halo   Population \\
GJ 581                    & M3V       & 9.5 {[}$+1.5, -1.5${]}      & 148.1 {[}$+0.9, -0.9${]}    & Thick Disk/Halo   Population \\
LHS 1802                  & M5V       & 9.97 {[}$+1.98, -2.39${]}   & 150.4 {[}$+2.4, -2.4${]}    & WD comp (LHS 1801)           \\
LHS 182                   & usdM0     & 11.5 {[}$+1.0, -1.5${]}     & 161.4 {[}$+5.2, -5.2${]}    & Halo Population              \\
GJ 1062                   & sdM2.5    & 11.5 {[}$+1.0, -1.5${]}     & 159.5 {[}$+2.1, -3.4${]}    & Halo Population              \\
LHS 3382                  & esdM2.5   & 11.5 {[}$+1.0, -1.5${]}     & 169.2 {[}$+9.3, -3.7${]}    & Halo Population              \\
LHS 216                   & esdM3$^{8}$   & 11.5 {[}$+1.0, -1.5${]}   & 174.4 {[}$+2.8,   -2.8${]}  & Halo Population              \\
Kapteyn's Star   (GJ 191) & sdM1.5$^{8}$ & 11.5 {[}$+1.0, -1.5${]}  & 153.2 {[}$+3.7, -3.7${]}    & Halo Population               \\
\enddata
\tablecomments{\footnotesize{$^1$Depending on the stellar parameter used, rotation periods in these clusters were measured for
objects as late as either M6 or M6.5.   $^2$\citet{2021ApJS..257...46G}   $^{3}$\citet{2022ApJ...931...45N}   $^{4}$\citet{2018ApJ...862...33A}, with additional $^{6}$rotation periods measured by us using \textit{ZTF} photometry
$^{5a}$Full Name: Gaia DR3 5172481276951287552    $^{5b}$Full Name: Gaia DR3 5172481203936294784}  $^{7}$\citet{2022ApJ...938..118D}    $^{8}$Classified using \citet{1997AJ....113..806G} and \citet{2019ApJS..240...31Z}}
\end{deluxetable*}
\end{longrotatetable}

\begin{deluxetable*}{llllll}
\tablecaption{Parameters Used to Derive Ages for the White Dwarf Companions \label{table:wdinfo}}
\tablehead{\colhead{WD Name}  & \colhead{WD Type}  & \colhead{Model}  & \colhead{$\log~g$}  & \colhead{$T_{\rm{eff}}$}  &  \colhead{Source(s)}}
\startdata
LP 856-53 & DA5   & H Thick & 8.06 {[}$+0.05,   -0.05${]}    & 9869 {[}$+92,   -92${]}  & $^6$, $^{17}$, $^{18}$, $^{19}$             \\
GJ 2131 B & DA3.9 & H Thick & 7.98 {[}$+0.04,   -0.04${]}    & 12573 {[}$+616,   -616${]} & $^6$, $^9$, $^{17}$                  \\
G 111-71  & DA6.5 & H Thick & 8.04 {[}$+0.02,   -0.02${]}    & 7592 {[}$+35, -35${]}      & $^4$, $^6$, $^9$, $^{18}$               \\
LP 775-53 & DA    & H Thick & 8.16 {[}$+0.05,   -0.05${]}    & 6509 {[}$+92, -92${]}    & $^6$, $^{18}$                     \\
EGGR 92   & DA4   & H Thick & 8.01 {[}$+0.04,   -0.04${]}    & 10590 {[}$+65,   -65${]} & $^3$, $^6$, $^{17}$, $^{18}$, $^{21}$                  \\
LP 587-53 & DA8.6 & H Thick & 7.97 {[}$+0.02,   -0.02${]}    & 5792 {[}$+48, -48${]}      & $^3$, $^4$, $^6$                   \\
LP 552-49 & DC    & H Thin  & 7.72 {[}$+0.4,   -0.4${]}    & 4431 {[}$+145,   -145${]}  & $^4$, $^{13}$                     \\
NSV 11920 & DBZ5  & H Thin  & 8.105 {[}$+0.02,   -0.02${]}    & 11070 {[}$+96,   -96${]}  & $^2$, $^7$, $^{11}$,$^{16}$, $^{28}$, $^{29}$       \\
Gaia DR3 4784 & non-DA & H Thin  & 8.15 {[}$+0.05,   -0.05${]}    & 9960 {[}$+112,   -112${]} & $^{28}$, $^{29}$                   \\
G 148-7   & DA3.1 & H Thick & 8.02 {[}$+0.02,   -0.02${]}    & 15964 {[}$+104,   -104${]} & $^3$, $^6$, $^9$, $^{13}$, $^{18}$, $^{26}$, $^{27}$ \\
LP 783-3    & DZ6.5 & H Thin & 8.10 {[}$+0.03,   -0.03${]}    & 7924 {[}$+97, -97${]}      & $^6$, $^{11}$, $^{16}$, $^{21}$       \\
G 130-5   & DA4   & H Thick & 8.01 {[}$+0.01,   -0.01${]}    & 12682 {[}$+58,   -58${]} & $^3$, $^6$, $^9$, $^{13}$, $^{18}$           \\
LP 498-26 & DB3   & H Thin  & 8.01 {[}$+0.04,   -0.04${]}    & 15868 {[}$+257,   -257${]} & $^5$, $^6$, $^7$                   \\
GJ 515    & DA4   & H Thick & 7.96 {[}$+0.03,   -0.03${]}    & 14346 {[}$+258,   -258${]} & $^9$, $^{17}$, $^{19}$, $^{21}$, $^{28}$     \\
40 Eri B  & DA2.9 & H Thin  & 7.94 {[}$+0.04,   -0.04${]}    & 17043 {[}$+453,   -453${]} & $^8$, $^9$, $^{20}$                  \\
G 201-39  & DA5.6 & H Thick & 7.95 {[}$+0.03, -0.03${]}      & 9002 {[}$+61, -61${]}      & $^3$, $^6$, $^{10}$, $^{17}$, $^{18}$          \\
G 87-29   & DQ8   & H Thin  & 8.00 {[}$+0.03,   -0.03${]}    & 6802 {[}$+79,   -79${]}  & $^4$, $^6$, $^{13}$, $^{24}$, $^{25}$          \\
LHS 230   & DA+DA & H Thick & 8.11 {[}$+0.01,   -0.01${]}    & 4955 {[}$+63,   -63${]}  & $^{12}$, $^{13}$, $^{21}$                      \\
LHS 1801  & DA    & H Thick & 7.926 {[}$+0.01,   -0.01${]}    & 5146 {[}$+22, -22${]}      & $^4$, $^6$, $^{13}$                  \\
L 266-196 & DA4.3 & H Thick & 8.05 {[}$+0.03,   -0.03${]}    & 10503 {[}$+442, -442${]}      & $^{17}$, $^{28}$, $^{29}$          \\
LP 906-23 & DA7   & H Thick & 7.985 {[}$+0.09,   -0.09${]}    & 6731 {[}$+205, -205${]}      & $^{28}$, $^{29}$                   \\
2MASS J23095848+5506491 & DA8.8 & H Thick & 8.116 {[}$+0.15,   -0.15${]}    & 5552 {[}$+28, -28${]}      & $^1$, $^{7}$, $^{13}$, $^{28}$\\
\enddata
\tablecomments{\footnotesize{$^1$\citet{2020MNRAS.499.1890M}  $^2$\citet{2020ApJ...900....2K}  $^3$\citet{2020ApJ...898...84K}  $^4$\citet{2019ApJ...878...63B}  $^5$\citet{2019RAA....19...88K}  $^6$\citet{2019MNRAS.482.4570G}  $^7$\citet{2018ApJ...857...56R}  $^8$\citet{2017ApJ...848...16B}  $^9$\citet{2017ApJ...848...11B}  $^{10}$\citet{2017MNRAS.469.2102A}  $^{11}$\citet{2017AJ....154...32S}  $^{12}$\citet{2016MNRAS.462.2295H}  $^{13}$\citet{2015ApJS..219...19L}  $^{14}$\citet{2014ApJ...796..128G}  $^{15}$\citet{2012MNRAS.425.1394K}  $^{16}$\citet{2012ApJS..199...29G}  $^{17}$\citet{2011ApJ...743..138G}  $^{18}$\citet{2011AA...531A...7G}  $^{19}$\citet{2009AA...505..441K}  $^{20}$\citet{2008AJ....135.1225H}  $^{21}$\citet{2008AJ....135.1239H}  $^{22}$\citet{2006AJ....132.1221H}  $^{23}$\citet{2006AA...450..331C}  $^{24}$\citet{2005ApJ...627..404D}  $^{25}$\citet{2001ApJS..133..413B}  $^{26}$\citet{1995ApJ...444..810B}  $^{27}$\citet{1992ApJ...394..228B}  $^{28}$\citet{2021MNRAS.508.3877G}  $^{29}$\citet{2023MNRAS.518.5106J}
}}
\end{deluxetable*}

\afterpage{
\startlongtable
\begin{deluxetable*}{lcccccccc}
\tablecaption{Rotation-based Age Determinations for Exoplanet-Hosting M Dwarfs\label{table:host_ages}}
\tablehead{
\colhead{Star Name}  & \colhead{$P_{\rm rot}$}  & \colhead{$P_{\rm rot}$ err$^1$}& \colhead{Age (Gyr)}  & \colhead{err$^1$}  & \colhead{Age (Gyr)}  & \colhead{err$^1$}  & \colhead{Relationship} & \colhead{Note$^2$} \\
\colhead{}  & \colhead{}  & \colhead{}& \colhead{Segmented Fit}  & \colhead{}  & \colhead{ODR Fit}  & \colhead{}  & \colhead{Used} & \colhead{}}
\startdata
USco1621 A              & 2.06  & 0.02 & 0.17          & 0.01 & --      & --   & mid-late     &      \\
HATS-74 A               & 4.75  & 0.05 & 0.18          & 0.02 & --      & --   & early        &      \\
AU Mic                  & 4.86  & 0.01 & 0.18          & 0.02 & --      & --   & early        &      \\
COCONUTS-2 A            & 2.83  & 0.28 & 0.19          & 0.01 & --      & --   & mid-late     &      \\
USco1556 A              & 4.67  & 0.05 & 0.24          & 0.01 & --      & --   & mid-late     &      \\
K2-284                  & 8.88  & 0.4  & 0.32          & 0.04 & --      & --   & early        &      \\
TOI-620                 & 8.99  & 0.09 & 0.33          & 0.03 & --      & --   & early        &      \\
K2-104                  & 9.3   & 0.4  & 0.34          & 0.04 & --      & --   & early        &      \\
K2-240                  & 10.8  & 0.1  & 0.42          & 0.05 & --      & --   & early        &      \\
Kepler-1512             & 10    & 0.01 & 0.47          & 0.02 & --      & --   & mid-late     &      \\
K2-141                  & 14.03 & 0.09 & 0.67          & 0.08 & --      & --   & early        &      \\
Kepler-1410             & 14.09 & 0.7  & 0.68          & 0.11 & --      & --   & early        &      \\
TOI-540                 & 0.73  & 0.01 & 0.72          & 0.05 & --      & --   & M4+          &      \\
EPIC 211822797          & 14.6  & 1.1  & 0.73          & 0.14 & --      & --   & early        &      \\
TRAPPIST-1              & 1.4   & 0.05 & 0.75          & 0.05 & --      & --   & M4+          &      \\
TOI-1227                & 1.65  & 0.04 & 0.76          & 0.05 & --      & --   & M4+          &      \\
2MASS J04372171+2651014 & 1.84  & 0.02 & 0.77          & 0.05 & --      & --   & M4+          &      \\
K2-25                   & 1.88  & 0.01 & 0.77          & 0.05 & --      & --   & M4+          &      \\
GJ 9066                 & 1.96  & 0.02 & 0.77          & 0.05 & --      & --   & M4+          &      \\
GJ 463                  & 14    & 0.1  & 0.79          & 0.05 & --      & --   & mid-late     &      \\
HATS-76                 & 15.16 & 0.2  & 0.79          & 0.10 & --      & --   & early        &      \\
Kepler-45               & 15.8  & 0.2  & 0.87          & 0.11 & --      & --   & early        &      \\
K2-415                  & 4.3   & 0.06 & 0.88          & 0.06 & --      & --   & M4+          &      \\
GJ 338 B                & 16.61 & 0.04 & 0.97          & 0.12 & --      & --   & early        & $*$  \\
Kepler-1229             & 17.63 & 0.88 & 1.13          & 0.21 & --      & --   & early        &      \\
GJ 685                  & 18.15 & 0.15 & 1.21          & 0.16 & --      & --   & early        &      \\
Kepler-1455             & 18.32 & 0.92 & 1.24          & 0.23 & --      & --   & early        &      \\
K2-345                  & 18.47 & 1.6  & 1.27          & 0.34 & --      & --   & early        &      \\
Gl 49                   & 18.86 & 0.1  & 1.34          & 0.18 & 2.34    & 0.06 & early        & $*$  \\
TOI-1685                & 18.66 & 0.71 & 1.44          & 0.16 & 2.78    & 0.08 & mid-late     &      \\
Kepler-395              & 19.92 & 1    & 1.56          & 0.31 & 2.39    & 0.08 & early        &      \\
Kepler-705              & 20.09 & 1    & 1.60          & 0.32 & 2.40    & 0.08 & early        &      \\
TOI-1201                & 21    & 2    & 1.94          & 0.52 & 2.84    & 0.10 & mid-late     &      \\
GJ 3470                 & 21.54 & 0.49 & 1.97          & 0.32 & 2.48    & 0.07 & early        &      \\
K2-264                  & 22.8  & 0.6  & 2.36          & 0.42 & 2.55    & 0.08 & early        &      \\
TOI-3714                & 23.3  & 0.3  & 2.54          & 0.41 & 2.58    & 0.07 & early        &      \\
K2-286                  & 23.8  & 3.7  & 2.63          & 0.53 & 2.61    & 0.22 & early        &      \\
BD-11 4672              & 25    & 0.3  & 2.69          & 0.51 & 2.67    & 0.08 & early        &      \\
Kepler-155              & 26.43 & 1.32 & 2.78          & 0.55 & 2.76    & 0.11 & early        &      \\
K2-95                   & 23.9  & 2.4  & 2.83          & 0.91 & 2.92    & 0.11 & mid-late     &      \\
GJ 514                  & 28    & 2.9  & 2.87          & 0.61 & 2.85    & 0.20 & early        &      \\
GJ 9827                 & 28.72 & 0.19 & 2.92          & 0.60 & 2.90    & 0.08 & early        &      \\
GJ 96                   & 29.6  & 2.8  & 2.97          & 0.65 & 2.96    & 0.20 & early        &      \\
L 168-9                 & 29.8  & 1.3  & 2.99          & 0.64 & 2.97    & 0.12 & early        &      \\
HD 147379               & 31    & 20   & 3.07          & 1.48 & 3.05    & 1.33 & early        &      \\
Kepler-1652             & 31.18 & 1.56 & 3.08          & 0.68 & 3.06    & 0.14 & early        &      \\
G 9-40                  & 29.85 & 1.01 & 3.09          & 0.34 & 3.08    & 0.10 & mid-late     &      \\
K2-332                  & 31.71 & 3.6  & 3.15          & 0.37 & 3.14    & 0.14 & mid-late     &      \\
LP 714-47               & 33    & 3    & 3.20          & 0.76 & 3.18    & 0.23 & early        &      \\
GJ 393                  & 34.15 & 0.22 & 3.28          & 0.78 & 3.26    & 0.10 & early        &      \\
TOI-776                 & 34.4  & 1.4  & 3.30          & 0.79 & 3.28    & 0.14 & early        &      \\
K2-155                  & 34.8  & 8.2  & 3.33          & 0.99 & 3.31    & 0.60 & early        &      \\
HATS-75                 & 35.04 & 0    & 3.35          & 0.81 & 3.33    & 0.10 & early        &      \\
K2-18                   & 38.6  & 0.5  & 3.35          & 0.45 & 3.34    & 0.10 & mid-late     &      \\
GJ 740                  & 35.56 & 0.07 & 3.38          & 0.83 & 3.37    & 0.10 & early        &      \\
TOI-1759                & 35.65 & 0.17 & 3.39          & 0.83 & 3.37    & 0.10 & early        &      \\
GJ 720 A                & 36.05 & 1.38 & 3.42          & 0.85 & 3.40    & 0.14 & early        &      \\
GJ 3293                 & 41    & 0.4  & 3.43          & 0.49 & 3.42    & 0.11 & mid-late     &      \\
HATS-71                 & 41.72 & 0.14 & 3.45          & 0.50 & 3.44    & 0.11 & mid-late     &      \\
LSPM J2116+0234         & 42    & 2    & 3.46          & 0.51 & 3.45    & 0.12 & mid-late     &      \\
TOI-1468                & 42.5  & 1.5  & 3.47          & 0.51 & 3.46    & 0.12 & mid-late     &      \\
KOI-4777                & 44    & 1    & 3.52          & 0.54 & 3.51    & 0.11 & mid-late     &      \\
HD 260655               & 37.5  & 0.4  & 3.53          & 0.91 & 3.51    & 0.11 & early        &      \\
Gl 686                  & 38.4  & 1.6  & 3.60          & 0.95 & 3.58    & 0.17 & early        &      \\
Kepler-235              & 39.49 & 1.97 & 3.68          & 1.00 & 3.67    & 0.19 & early        &      \\
GJ 9689                 & 39.97 & 0.39 & 3.72          & 1.01 & 3.70    & 0.12 & early        &      \\
TYC 2187-512-1          & 40    & 1    & 3.72          & 1.02 & 3.71    & 0.14 & early        &      \\
K2-3                    & 40    & 2    & 3.72          & 1.03 & 3.71    & 0.20 & early        &      \\
Kepler-560              & 50.47 & 2.52 & 3.74          & 0.66 & 3.73    & 0.15 & mid-late     &      \\
TOI-674                 & 52    & 5    & 3.79          & 0.70 & 3.78    & 0.21 & mid-late     &      \\
TOI-700                 & 54    & 0.8  & 3.86          & 0.72 & 3.85    & 0.13 & mid-late     &      \\
GJ 367                  & 58    & 6.9  & 4.01          & 0.85 & 4.00    & 0.29 & mid-late     &      \\
TOI-1235                & 44.7  & 4.5  & 4.12          & 1.31 & 4.11    & 0.42 & early        &      \\
GJ 536                  & 45.39 & 1.33 & 4.18          & 1.30 & 4.17    & 0.18 & early        &      \\
GJ 1252                 & 64    & 4    & 4.23          & 0.96 & 4.22    & 0.21 & mid-late     &      \\
LHS 1678                & 64    & 22   & 4.23          & 1.28 & 4.22    & 0.87 & mid-late     &      \\
HD 180617               & 46.04 & 0.2  & 4.24          & 1.33 & 4.23    & 0.14 & early        &      \\
GJ 4276                 & 64.3  & 1.2  & 4.25          & 0.95 & 4.24    & 0.15 & mid-late     &      \\
TOI-1695                & 47.7  & 2.2  & 4.40          & 1.44 & 4.38    & 0.25 & early        &      \\
LHS 1815                & 47.8  & 0.7  & 4.40          & 1.43 & 4.39    & 0.16 & early        &      \\
GJ 1265                 & 70    & 0.7  & 4.47          & 1.10 & 4.46    & 0.15 & mid-late     &      \\
GJ 436                  & 71.4  & 0.4  & 4.53          & 1.14 & 4.52    & 0.15 & mid-late     & $*$  \\
TOI-122                 & 72    & 22   & 4.56          & 1.48 & 4.55    & 0.94 & mid-late     &      \\
L 98-59                 & 80.9  & 5    & 4.94          & 1.44 & 4.94    & 0.28 & mid-late     &      \\
TOI-2136                & 82.56 & 0.45 & 5.02          & 1.48 & 5.01    & 0.17 & mid-late     &      \\
YZ Cet                  & 83    & 15   & 5.04          & 1.64 & 5.03    & 0.72 & mid-late     & $*$  \\
Teegarden's Star        & 86.3  & 1.3  & 5.20          & 1.60 & 5.19    & 0.19 & mid-late     & $*$  \\
Proxima Cen             & 86.4  & 2.5  & 5.20          & 1.61 & 5.19    & 0.22 & mid-late     & $*$  \\
GJ 3512                 & 87    & 5    & 5.23          & 1.64 & 5.22    & 0.30 & mid-late     &      \\
GJ 411                  & 56.15 & 0.27 & 5.27          & 2.03 & 5.27    & 0.18 & early        & $*$  \\
GJ 3323                 & 88.5  & 0.89 & 5.30          & 1.68 & 5.30    & 0.19 & mid-late     & $*$  \\
GJ 3779                 & 95    & 5    & 5.63          & 1.94 & 5.62    & 0.33 & mid-late     &      \\
G 264-012               & 100   & 6    & 5.89          & 2.16 & 5.89    & 0.39 & mid-late     &      \\
TOI-237                 & 102   & 22   & 6.00          & 2.53 & 6.00    & 1.24 & mid-late     &      \\
LTT 3780                & 104   & 15   & 6.11          & 2.46 & 6.11    & 0.88 & mid-late     &      \\
Wolf 1061 (GJ 628)      & 108.7 & 1.5  & 6.38          & 2.53 & 6.38    & 0.26 & mid-late     & $*$  \\
GJ 3929                 & 122   & 13   & 7.21          & 3.34 & 7.22    & 0.91 & mid-late     &      \\
GJ 251                  & 122.1 & 1.9  & 7.22          & 3.24 & 7.22    & 0.32 & mid-late     &      \\
GJ 1132                 & 122.3 & 6    & 7.23          & 3.27 & 7.24    & 0.50 & mid-late     &      \\
Ross 128                & 123   & 1.2  & 7.28          & 3.29 & 7.28    & 0.30 & mid-late     &      \\
GJ 1214                 & 124.7 & 5    & 7.40          & 3.41 & 7.40    & 0.45 & mid-late     &      \\
GJ 1002                 & 126   & 15   & 7.48          & 3.62 & 7.49    & 1.08 & mid-late     &      \\
CD Cet                  & 126.2 & 1.3  & 7.50          & 3.48 & 7.50    & 0.32 & mid-late     &      \\
GJ 486                  & 130.1 & 1.6  & 7.77          & 3.73 & 7.78    & 0.34 & mid-late     &      \\
LHS 1140                & 131   & 1.3  & 7.84          & 3.79 & 7.84    & 0.34 & mid-late     &      \\
TOI-1634                & 77    & 26   & 8.27          & 6.44 & 8.29    & 4.70 & early        &      \\
GJ 625                  & 77.8  & 5.5  & 8.41          & 4.70 & 8.43    & 1.07 & early        & $*$  \\
GJ 1151                 & 140   & 10   & 8.51          & 4.49 & 8.52    & 0.87 & mid-late     &      \\
Wolf 1069               & 160   & 10   & 10.23         & 6.19 & 10.25   & 1.06 & mid-late     &      \\
GJ 273                  & 160.8 & 2.5  & 10.31         & 6.20 & 10.33   & 0.54 & mid-late     & $*$  \\
GJ 3473                 & 168.3 & 4.2  & 11.04         & 6.98 & 11.07   & 0.68 & mid-late     &      \\
HD 238090               & 96.7  & 3.7  & 12.64         & 8.81 & 12.72   & 1.19 & early        &      \\
\enddata
\tablecomments{\footnotesize{$^1$As explained in the text, uncertainties along the younger branch
(Age $\lesssim 2.9$ Gyr) do not account for the full scatter of rotation rates in clusters, and
are therefor underestimated. Additionally, some rotation rates from the Exoplanet Archive did not
have uncertainties.
$^2$ An asterisk in the Note column indicates that the rotation rate was updated to match the value presented either
elsewhere in this paper, or in the follow-up paper.
}}
\end{deluxetable*}
}


\vspace{10mm}

We would like to acknowledge the tireless work of the \textit{RCT Consortium} members, 
including former members David Laney and Louis-Gregory Strolger, for the operation
and maintenance of the telescope and equipment, without which this study would
not have been possible. We also wish to acknowledge the builders and operators of
the \textit{Skynet} telescope network and the \textit{Zwicky Transient Facility}. 
Partial support for this project was provided by CXO grant GO0-21020X.

SGE thanks Louis Amard for invaluable discussions and access to his M dwarf interior
models. SGE also thanks Andrej Pr{\v{s}}a for very helpful discussions regarding the
regression fitting.

%

\vspace{5mm}
\facilities{KPNO:RCT, CTIO:PROMPT}


\software{astropy \citep{2013AA...558A..33A,2018AJ....156..123A,2022ApJ...935..167A},  
          Matplotlib \citep{Hunter:2007},
          Pandas \citep{mckinney-proc-scipy-2010},
          NumPy \citep{harris2020array},
          SciPy \citep{2020SciPy-NMeth},
          AstroImageJ \citep{2017AJ....153...77C}
          }

\end{document}